\documentstyle[12pt,epsf]{article}
\newcommand{\nc}{\newcommand}
\def\frac#1#2{{\textstyle {#1 \over #2}}}

\def\d{{\rm d}}

\nc{\beq}{\begin{equation}}
\nc{\eeq}{\end{equation}}
\nc{\beqa}{\begin{eqnarray}}
\nc{\eeqa}{\end{eqnarray}}
\nc{\lsim}{\begin{array}{c}\,\sim\vspace{-21pt}\\< \end{array}}
\nc{\gsim}{\begin{array}{c}\sim\vspace{-21pt}\\> \end{array}}
\nc{\appA}{}
\nc{\appB}{}
\nc{\appC}{}
\nc{\appD}{}
\nc{\appE}{}
\nc{\Eqr}[1]{(\ref{#1})}
\nc{\mysection}[1]{\setcounter{equation}{0}\section{#1}}

\nc{\myappendix}[1]{\section*{#1}\setcounter{equation}{0}}

\def\bra{\langle }
\def\ket{\rangle }

\def\dk{d\bar{k}}
\def\dr{d\bar{r}}
\def\dq{d\bar{q}}
\def\bPsi{\bar\Psi}
\def\bpsi{\bar\psi}

\def\bta{\bar\eta}

\def\&{and}

\def\DS {D\!\!\!\!/}
\def\dS{\partial\!\!\!/}
\def\kS{k\!\!\!/}
\def\qS{q\!\!\!/}
\def\pS{p\!\!\!/}
\def\rS{r\!\!\!/}
\def\d{\partial_\mu} 
\def\A { A_\mu (x) }
\def\AS {A\!\!\!/}

\def\C {C}

\def\bu{ \bar{u} }

\def\D{ {\cal D} }
 
\def\ap#1#2#3{           {\it Ann. Phys. (NY) }{\bf #1}, #2 (19#3)}

\def\com#1#2#3{          {\it Comm. Math. Phys. }{\bf #1}, #2 (19#3)}

\def\nc#1#2#3{           {\it Nuovo Cim.  }{\bf #1}, #2 (19#3)}
\def\ncl#1#2#3{          {\it Nuovo Cim. Lett.  }{\bf #1}, #2 (19#3)}
\def\np#1#2#3{           {\it Nucl. Phys. }{\bf #1}, #2 (19#3)}
\def\pl#1#2#3{           {\it Phys. Lett. }{\bf #1}, #2 (19#3)}
\def\p#1#2#3{           {\it Physics }{\bf #1}, #2 (19#3)}
\def\pr#1#2#3{           {\it Phys. Rev. }{\bf #1}, #2 (19#3)}
\def\prep#1#2#3{         {\it Phys. Rep. }{\bf #1}, #2 (19#3)}
\def\prl#1#2#3{          {\it Phys. Rev. Lett. }{\bf #1}, #2 (19#3)}

\def\proc#1#2#3{          {\it Proc. R. Soc. London }{\bf #1}, #2 (19#3)}
\def\rmp#1#2#3{          {\it Rev. Mod. Phys. }{\bf #1}, #2 (19#3)}

\def\ra#1#2#3{            {\bf #1}, #2 (19#3)(E)}

\begin{document}
\begin{titlepage}
\renewcommand{\thefootnote}{\fnsymbol{footnote}}
\begin{center}
\hfill
\vskip 1 cm
{\large \bf Continuous non-perturbative regularization of QED}
\vskip 1 cm
{
  {\bf J. L. Jacquot}\footnote{jacquot@lpt1.u-strasbg.fr}
   \vskip 0.3 cm
   {\it Laboratoire de Physique Th\'eorique,
        3 rue de l'Universit\'e,
        67084 Strasbourg Cedex, FRANCE}\\ }
  \vskip 0.3 cm
\end{center}

\vskip .5 in
\begin{abstract}
We regularize in a continuous manner the path integral of QED by construction 
of a 
non-local version of its action by means of a regularized form of Dirac's 
$\delta$ 
functions.
Since the action and the measure are both invariant under the gauge group, this 
regularization scheme is intrinsically non-perturbative.
Despite the fact that the non-local action converges formally to the 
local 
one as the cutoff goes to infinity, the regularized theory keeps trace of the 
non-locality  through the  appearance of a quadratic divergence in the 
transverse part 
of the polarization operator.
This term which is uniquely defined by the choice of the cutoff functions can be 
removed by a redefinition of the regularized action.
We notice that as for chiral fermions on the lattice, there is an obstruction to 
construct a  continuous and non ambiguous regularization in four 
dimensions.  
With the help of the regularized equations of motion, we calculate the one 
particle 
irreducible 
functions which are known to be divergent by naive power counting at the one 
loop 
order.
\end{abstract}
\end{titlepage}
\renewcommand{\thepage}{\arabic{page}}
\setcounter{page}{1}
\setcounter{footnote}{0}
\mysection{Introduction}
As it is well known gauge invariant quantum field theory is a powerful tool to 
describe 
fundamental 
interactions in physics.
However these theories suffer from ultraviolet (UV) divergences if we impose 
the locality of 
the 
interactions.
Over the past decade since the birth of QED great efforts have been made in 
order to 
regularize these 
theories in a gauge invariant manner in order to preserve the Ward-Takahashi 
(WT) 
identities, since it appears hopeless to cure this problem by the construction 
of 
non 
local 
interactions \cite{PE}.
This is a necessity dictated by the renormalization program.
There are essentially two categories of gauge invariant regularization schemes, 
a discrete one and a 
continuous one.
In the first category the path integral of the theory is discretized on a 
lattice.
This allows to make non-perturbative, but mostly non analytical predictions.
On the other hand the second category of continuous gauge invariant 
regularization schemes is intrinsically perturbative, in the sense that in 
momentum 
space these schemes basically 
regularize  
the 
Feynman integrals or the lowest order Green's functions in configuration space.
This is the case for Pauli-Villars regularization \cite{PV}, dimensional 
regularization \cite{DR,CO}, 
Schwinger's 
proper time 
methods inspired regularization \cite{SR}, $\zeta $ function 
regularization \cite{ZR}, differential 
regularization \cite{DDR}
and Fujikawa's 
type of regularization \cite{FR,FFR}.

If one wants to study the non-perturbative properties or the different phases 
of 
a gauge 
theory as a function of the scale, a general method is  the Wilson-Kadanoff 
\cite{WK} 
approach.
In this case it is more sensible to work, in four dimensions, with a continuous 
regularized 
and gauge 
invariant version of the path integral of the theory.  
In recent years, some progress have been made in this direction.
For instance, gauge invariant regularization of chiral gauge theories can be 
achieved by a 
generalization 
of the Pauli-Villars method \cite{PPR,FFR} i.e. by the addition of a finite or 
infinite 
set 
of 
non physical 
regulator 
fields at the lagrangian level.
In the same spirit a generalization of Schwinger's proper time methods, the 
operator cutoff 
regularization \cite{CR}, allows 
to calculate effective gauge invariant regularized action at the one loop 
level.

In the present work we follow Wilson' s idea, but we regularize 
non-perturbatively the 
path integral 
of 
QED in a continuous manner.
This is done by a suited smearing of the point-like interactions of the fields 
by cutoff 
functions which are a regularized form of Dirac's $\delta $ function.
We choose the cutoff functions to act only in the (UV) domain and not in the 
infrared (IR) one.
In this sense, for finite (UV) cutoff our action is non-local, but converges 
asymptotically to 
the 
local one as the cutoff goes to infinity.
We should notice that an earlier attempt was made in constructing a non-local 
regularization 
of 
gauge 
invariant theory \cite{NR}.
In this regularization scheme only the action is invariant under a non-local 
transformation\footnote{This transformation is not the representation of a 
group.} which 
plays the role of the gauge transformation.
The invariance of the path integral is maintained by adding a term in the 
action 
which cancels the contribution of the transformed measure.
Since this term must be worked out order by order in perturbation theory, this 
method 
belongs to the category of perturbative regularization schemes.  
 
In our case, both the cutoff action and the measure are invariant under the 
gauge 
group, therefore there is no need 
to define the regularized partition function perturbatively through the 
regularization 
of the Feynman 
diagrams.
As a result the path integral is regularized as a whole in a gauge invariant 
manner.
This feature gives us the opportunity to apply directly non-perturbative 
techniques to 
the 
regularized partition function in order to derive the physical properties of 
the 
theory. 
For instance, in addition to numerical simulations, we can study the variation 
of the 
path 
integral under an 
infinitesimal variation of the cutoff scale in order to derive Polchinski's 
\cite{POLCH} 
like 
exact renormalization group equation.
We can also derive the 
regularized equations of motion, i.e. the regularized Dyson-Schwinger (DS) 
equations, directly from the regularized partition function of the 
theory. 
Since this scheme keeps track of the dimension of any regularized integral 
through the 
logarithm or the powers of the cutoff, the domain of validity of the 
regularized (DS) 
equations is not only restricted to the (UV) region, but can be extended to all 
range of 
the cutoff scale in contradistinction to dimensional regularization \cite{DR}.
Thus this regularization scheme can give an 
insight into non-perturbatives properties of the theory, even in the context of 
perturbative calculations where the (DS) equations 
appear 
as a resummation of perturbative series.
For example this scheme is suitable for the calculation of the vacuum 
expectation value of 
operators like the condensates.
Moreover, as this method is continuous and works strictly in four dimensions, 
it can be 
applied to the treatment of the axial anomaly which is known to connect both 
the (UV) 
and the (IR) regions.
 
In order to show that the partition function of (QED) is indeed finite, we 
derive first the regularized Dyson-Schwinger (DS) equations which are the 
basic 
ingredients for the skeleton expansion.
The class of regularization is imposed by the choice of a cutoff function.
We choose to mimic Schwinger's proper time regularization.
Then, despite the fact that our regularization is non-perturbative, we deduce 
the 
regularized form of the relevant one particle irreducible (1PI) 
functions 
which are 
known 
to 
be potentially divergent by power counting\footnote{The direct study of the 
regularized 
partition function with group renormalization 
techniques will be done in a forthcoming paper.} perturbatively from the 
regularized 
(DS) 
equations.
We show that we recover the known result for the mass operator.
This is also true for the vertex function if we assume that the electron 
lines are on 
mass-shell.
Since the regularization of the path integral is gauge invariant by 
construction,  the polarization operator is transverse and regularized.
The finite part of the polarization operator has the standard form, the only 
discrepancy 
occurs 
in its divergent part where a quadratic divergence appears. 
This divergent term can be removed by adding to the lagrangian a gauge 
invariant 
counterterm $ S_{\Lambda } $ which is 
quadratic in the photon field. 
The paper is organized as follows.
In section 2 we define the regularized gauge invariant action in configuration 
and in momentum 
space.
The regularized form of the (DS) equations and of the (WT) identity are given 
too.
We derive the relevant (1PI) functions from the equations of motion in the next 
sections.
Section 3 is devoted to the polarization operator and to the 2n-photon 
amplitudes.
In addition to this section the role of $ S_{\Lambda } $ is clarified.
We calculate the electron mass operator and the vertex function in section 4.
\mysection{The regularized gauge invariant action, the (DS) \\
equations and the (WT) identity}
In order to regularize the action of QED  in four dimensions  we construct a 
non-local 
cutoff action which converges 
asymptotically to 
the 
standard non-regularized one as the cutoff tends to infinity.
This can be done by a suited smearing of the point-like interactions of the 
fields 
by 
  the scalar cutoff functions  
\beq
\label{cutf}
\rho_i (x,y) ~=~ \int \dk ~e^{-ik(x-y)}\rho_i (k,\Lambda ),  
\eeq
which are symmetrical and  are  regularized forms of 
Dirac's $\delta $ 
function, i.e.  
\beqa
\label{limcutf} 
\ & & \lim_{\Lambda \to \infty}\rho_i (k,\Lambda )~=~1, \\
\label {delta}
\ & & \lim_{\Lambda \to \infty}\rho_i (x,y)~=~\delta (x-y).  
\eeqa
Here and in the following we work in Minkowski space.
We choose the signature of the metric to be $ (1,-1,-1,-1) $ and the 
notation $ dx 
\equiv  d^4x $ and 
$ 
\dk 
\equiv  d^4k/(2\pi )^4 $.

In all analytically methods used to calculate or to define the Greens's functions 
of 
a 
given theory the inverse of the energy kinetic terms appear through the 
propagators.
This is the case for non-perturbative methods based on exact renormalization group 
equation or Dyson-Schwinger expansion based on the use of the equations of motion 
of 
the 
theory and or pure perturbative methods based on Feynman diagrams expansion.
It follows that the regularized action $S_{Reg}$  of QED must obey three conditions.

(1) $S_{Reg}$ must be gauge invariant.

(2) In any expression composed of product of (1PI) functions the 
inverse of the cutoff functions (\ref{cutf}) associated to the kinetic terms which 
occur in the propagators must not  cancel the cutoff 
functions  associated to the vertices.

(3) When the cutoff $\Lambda$ goes to infinity $S_{Reg}$ must converge to the local 
non-regularized action of QED.

\begin{figure}
	\epsfxsize=8cm
	\epsfysize=5.5cm
	\centerline{\epsffile{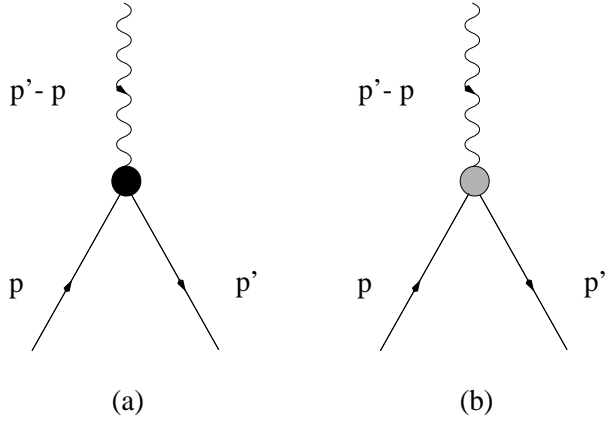}}
\caption{The vertex associated to the matrix $-e\Gamma_{\mu }(p,p')$ (a) and to its 
transversal part (b).\label{vert1}}
\end{figure}

Since in QED the photon has no self-interaction we need only to regularize  
the 
fermionic 
part of the action. 
A solution to these constraints is to regularize separately in a gauge 
invariant manner  the  kinetic, the interaction  and the mass terms.
The minimal solution is to keep the mass term unregularized.
Then the building blocks needed to construct $S_{Reg}$ are the smeared gauge field 
\beq
\label{sma}
\mathbf{A}^{\mu}(x)~=~\int dy~ \rho_3(x,y)A^{\mu }(y),
\eeq
the functional of the smeared gauge field
\beq
\label{L}
\ L(z,z') ~=~ \int dy~C_{\mu }(z,z',y)\mathbf{A}^{\mu}(y)
\eeq
and the functionals
\beq
\label{psixz}
\mathbf{\Psi}_i (x)~=~\int dy~\rho_i(x,y)e^{ieL(x,y )}\psi (y)
\eeq 
of the smeared fermion and gauge fields  which remembers Schwinger's point splitting 
\cite{SR}.
Likewise in lattice gauge theory the functional $L(z,z')$ plays 
the 
role 
of a link.
The form of the C-number function $C_{\mu }$ is imposed by the necessity of gauge 
covariance.
In order that  the functional $\mathbf{\Psi}_i (x)$ transforms covariantly  under 
the 
gauge transformation 
\beqa
\label {gautrans}
\ \A ~&\rightarrow &~\A -\frac{1}{e}\d \Lambda(x) \nonumber \\
\ \psi(x)~&\rightarrow &~e^{i\int dy~\rho_3(x,y)\Lambda(y)}\psi(x),
\eeqa
the variation of $L(z,z')$ must be
\beq
\label {deltaL}
\ \delta L(z,z')~=~\frac{1}{e} \int 
dy~\left(\rho_3(z,y)-\rho_3(z',y)\right)\Lambda (y).
\eeq
This imposes that the  divergence of  the C-number function $C_{\mu }$ verifies 
\beq
\label{dC} 
\ \partial^{\mu }C_{\mu }(z,z',x) ~=~ \delta(x-z)-\delta(x-z').
\eeq
In addition  the choice of $C_{\mu }$ is such that
\beq
\label{Ccond}  
\ L(z,z)~=~0.
\eeq
This is the necessary condition to recover the local fundamental fields from 
(\ref{sma}) and (\ref{psixz}) when the cutoff tends to infinity.
In fact due the properties (\ref{delta}) of the cutoff functions we have now,
\beqa
\label{limsma}
\ & & \lim_{\Lambda \to \infty} \mathbf{A}^{\mu}(x)~=~A^{\mu}(x) \nonumber \\
\ & & \lim_{\Lambda \to \infty}\mathbf{\Psi}_i (x)~=~\psi(x).
\eeqa
For $C_{\mu }$ we choose the less (UV) divergent solution
\beq
\label{C} 
\C_{\mu }(z,z',x) ~=~ i\int \dr ~\frac{r\mu }{r^2} \left( 
e^{-ir(x-z)}-e^{-ir(x-z')}\right).
\eeq 
Then the regularized action $S_{Reg}$ of QED is the sum of the fermionic term 
$S_{Fermion}$ and of 
the 
pure gauge term $S_{Gauge}$ which are defined by
\beqa
\label{actferm} 
\ S_{Fermion}  ~&=&~\int dxdzdz'~ \bpsi (z)~e^{ieL(z,x)}\Bigg [\rho_1 
(z,x)\bigg 
(i\dS ~+~e(\dS L(x,z'))\bigg )\rho_1(x,z' ) \nonumber  \\
& &-~e\rho_2 (z,x)\bigg( \mathbf{\AS} (x)~+~(\dS L(x,z')) \bigg) \rho_2(x,z' 
)\Bigg ] e^{ieL(x,z' )}\psi (z') \nonumber \\ 
& &-~m~\int dx~\bpsi\psi \\
\label{actgau}
\ S_{Gauge}  ~&=&~-\int dx \left[\frac {1}{4}F_{\mu \nu 
} F^{\mu \nu 
} ~+~\frac{1}{2\xi }\left(\partial^{\mu }A_{\mu } \right)^2 
\right]~+~S_{\Lambda}.   
\eeqa
The contribution $S_{\Lambda}$ of (\ref{actgau}) is gauge invariant and will 
play 
the role of a counterterm for the polarization operator.
This term will be discussed in more details in section 3.
Since $ L(z,z')$ is defined by (\ref{L}) and (\ref{C}), the partial derivative  $\dS 
L(x,z')$ which enters in  (\ref{actferm}) is  a function of $x$ only.
This implies that the kinetic and interaction terms of the regularized action  
(\ref{actferm}) have the general form $\int dx~ \mathbf{\bPsi}_i \mathbf{K}_i 
\mathbf{\Psi}_i$, where $\mathbf{K}_i$ is a functional of the smeared gauge field 
(\ref{sma}).
As a result one can readily check that each of the three terms of (\ref{actferm}), 
which 
correspond respectively to the kinetic, the interaction and the mass term,  
are by construction separately 
gauge 
invariant.
When the cutoff $\Lambda $ tends to infinity, it follows from the relations 
(\ref{limsma})  that the terms proportional to $\dS L(x,z')$ cancel and that  the 
regularized action $S_{Reg}-S_{\Lambda}$ converges formally to 
the 
standard 
local action 
of 
QED.
In deriving the relations (\ref{limsma}) and the asymptotic form of the regularized 
action we have assumed that we  can formally 
interchange the limit with the integral symbol.
This assumption is not exactly true\footnote{Even in one dimension this is not 
in 
general the 
case 
for improper integral.}, and will be to the origin of the appearance of a 
quadratic divergence 
in 
the 
polarization operator.
If we use   the commutation relation 
\beq
\label {com}
\ [~i\dS ~,~e^{ieL(x,z')}~]~=~-e~e^{ieL(x,z')}(\dS L(x,z')), 
\eeq
we can rewrite  (\ref{actferm}) in a more concise form as
\beqa
\label{actfer} 
\ S_{Fermion}  ~&=&~\int dxdzdz'~ \bpsi (z)~e^{ieL(z,z')}\Bigg [\rho_1 
(z,x) i\dS\ 
\! \rho_1(x,z' )  \nonumber  \\
& &-~e\rho_2 (z,x)\bigg ( \mathbf{\AS} (x)~+~(\dS L(x,z'))\bigg  
)\rho_2 
(x,z' 
)\Bigg ]\psi (z')\nonumber  \\
& &-~m~\int dx~\bpsi\psi. 
\eeqa
From now on, the regularized gauge invariant path integral which we consider in 
configuration space is
\beq
\label {zx}
\ Z(\eta,\bta,J)~=~\int \D\psi\D\bpsi\D A_{\mu}~e^{i\left(S(\psi ,\bpsi 
,A)~+~\int 
dx~(\bta(x)\psi(x)+\bpsi(x)\eta(x)+J_{\mu }(x)A^{\mu }(x))\right)}
\eeq
where $S(\psi ,\bpsi ,A)$ is the sum of (\ref{actgau}) and (\ref{actfer}).
The $ \eta,~\bta ~and~ J_{\mu } $ are respectively the sources for the fermion 
and gauge 
fields.
As before  one can notice that the new form (\ref{actfer}) of the fermionic 
part 
of 
the
action is still
composed 
of three terms which are separately gauge invariant.
The fact that the whole action $S(\psi ,\bpsi ,A)$ is a regularized action of 
QED in 
four dimensions will become transparent in 
momentum 
space  where 
we give the 
representation of the partition function (\ref{zx}) and of the Legendre 
transform of 
the (1PI) 
generating 
functional 
\beq  
\label {gf}
\ G(\eta,\bta,J) \equiv \log Z(\eta,\bta,J).  
\eeq
For the partition function we have
\beq
\label {zp}
\ Z(\eta,\bta,J)~=~\int \D\psi\D\bpsi\D A_{\mu}~e^{i\left(S(\psi ,\bpsi 
,A)~+~\int \dk ~ (\bta(k)\psi(k)+\bpsi(k)\eta(k)+J_{\mu }(-k)A^{\mu 
}(k))\right)}.
\eeq
In this expression  the action is 
\beqa
\label {skone}
\ S(\psi,\bpsi,A)~&=&~\int\dk 
d\bar{p} d\bar{p'}~\bpsi(p) K(k,p,p') \psi(p')~+~S_{Gauge} \\
\label {sgauge}
\ S_{Gauge}~&=&-~\frac{1}{2} 
\int \dk ~ \left( k^2 g_{\mu \nu } - k_{\mu } 
k_{\nu } (1-\frac{1}{\xi }) \right) A^{\mu }(k) A^{\nu }(-k) + S_{\Lambda}, 
\eeqa
and the kernel $K$ is given by
\beqa
\label{sktwo}
\ K(k,p,p')~&=&~F(p-k,k-p')\rho_1^2(k)\kS-m((2\pi)^4)^2 
\delta(p-k)\delta(p'-k)\nonumber \\& &-~e\int \dq 
~\rho_3(q) A^{\mu}(q) \Gamma_{\mu }(q)F(p-k,k-q-p')\rho_2(k)\rho_2(k-q),
\eeqa
where the transverse matrix  $\Gamma_{\mu }(q)$ \footnote{See the properties of this 
matrix in appendix A.} is defined by
\beq
\label {gam}
\ \Gamma_{\mu }(q)~~=~~\gamma_{\mu }-\qS\frac{q_{\mu }}{q^2},
\eeq
and $F(p,q)$ is the Fourier transform of $e^{ieL(x,z')}$.
With the help of the recursion formulas of appendix B which define  $F(p,q)$, 
we 
can 
rewrite the action (\ref{skone}) in the final form 
\beqa
\label {skthree}
\ S(\psi,\bpsi,A)~&=&~\int\dk ~\bpsi(k) \left(\rho_1^2(k)\kS-m 
\right)\psi(k)~-~e\int d\bar{p} d\bar{p'}~\bpsi(p)A^{\mu}(p-p')\Gamma_{\mu 
}(p,p')\psi(p')\nonumber \\
& &+~\sum_{n=0}^{+\infty} \frac{(ie)^{(n+2)}}{(n+2)!}\int\dk 
d\bar{p} d\bar{p'}~\bpsi(p)\bigg[F_{n+2}(p-k,k-p')\rho_1^2(k)\kS 
~+~i(n+2)\nonumber 
\\
& & \int 
\dq~\rho_3(q) A^{\mu}(q)\Gamma_{\mu }(q)F_{n+1}(p-k,k-q-p')\rho_2(k)\rho_2(k-q) 
\bigg]\psi(p')\nonumber \\
& &+~S_{Gauge}.
\eeqa
Here the matrix 
  $\Gamma_{\mu }(p,p')$ is defined in terms of (\ref{gam}) by
\beq
\label {Gamone}
\ \Gamma_{\mu}(p,p')~=~\rho_3(p-p')\left[\rho_2(p)\rho_2(p')\Gamma_{\mu}(p-p') 
+ 
\frac{(p-p')_{\mu}}{(p-p')^2}\left(\rho_1^2(p) \pS-\rho_1^2(p') 
\pS'\right) 
\right],
\eeq
and has the property
\beq
\label {Gamtwo}
\ \lim_{\Lambda \to \infty}\Gamma_{\mu}(p,p')~=~\gamma_{\mu}.
\eeq
Owing to the property (\ref{Gamtwo})  the first and the second term of  
(\ref{skthree}) converge to the standard fermionic part of the action of QED 
when 
the cutoff goes to infinity.
As for the third term, it vanishes identically due to the properties  
(\ref{recur2}) 
of 
the $F_n$.
This term which describes an infinite set of vertices  is needed to ensure the 
gauge invariance of the regularized amplitudes.
Therefore the whole action (\ref{skthree}) converges formally to the 
non-regularized action of QED.
Since the third term of (\ref{skthree}) vanishes asymptotically, we expect that the  
matrix  $-e\Gamma_{\mu }(p,p')$ (\ref{Gamone}) or its transversal part 
$-e\rho_3(p-p')\rho_2(p)\rho_2(p')\Gamma_{\mu}(p-p')$ associated to the vertices of 
Fig.\ref{vert1}   will play an essential role in the representation  of  amplitude by 
Feynman 
diagrams.
Notice that the matrix $\Gamma_{\mu }(p,p')$ (\ref{Gamone}) is unambiguously defined 
by the values of the momentas of the two electrons lines because $\Gamma_{\mu 
}(p,p')$ is symmetrical in the variables $p$ and $p'$.
 
From the action (\ref{skthree}) it is readily seen that the free electron and 
photon 
propagators behave respectively for high $k^2$ like $1/m$ and $1/k^2$.
Since at each vertex is associated a cutoff function, it follows that the path 
integral 
(\ref{zp}) defined by the action (\ref{skthree}) is  finite.
Apart from the fact that in euclidian space the (UV) cutoff functions must be 
rapid 
decreasing 
functions of the squared momenta and must be defined for all 
values 
of the 
momentum, the 
choice of their form is quite arbitrary.
For instance if all the cutoff functions (\ref{cutf})  are identical  the 
action  (\ref{skthree})  which is expressed in terms of the fundamental fields does 
not 
simplify.
However when expressed in configuration space in terms of the smeared fields 
(\ref{sma}) 
and (\ref{psixz}) the fermionic part of the regularized action take the compact form
\beq
\label{comactferm}
\ S_{Fermion}~=~ \int dx~\mathbf{\bPsi}i\mathbf{\DS}\mathbf{\Psi}-m\int 
dx~\bpsi\psi,
\eeq
where $i\mathbf{\DS}$ is the smeared covariant derivative defined by
\beq
\label{smearD}
i\mathbf{\DS}~=~i\dS-e\mathbf{\AS}.
\eeq
From a practical point of view it is better to leave the cutoff functions 
unconstrained, 
then in the following we consider the general action (\ref{skthree}).
In order to check that the theory is indeed finite, we shall derive  in the next 
sections 
the  
regularized 
(1PI) functions from the  regularized equations of motion. 

For this purpose we also work out  the Legendre transform of 
(\ref{gf}) in momentum space.
It is given in terms of the vacuum 
expectation 
value of the fields
\beq
\label {classfield}
\ \bpsi(k)~=~i(2\pi )^4 \frac {\delta G}{\delta \eta (k)}~~~~\psi(k)~=~-i(2\pi 
)^4 \frac 
{\delta 
G}{\delta 
\bta (k)}~~~~A_{\mu }(-k)~=~-i(2\pi )^4 \frac {\delta G}{\delta J^{\mu }(k)}  
\eeq
by
\beq
\label {gamma}
\ \Gamma (\psi,\bpsi,A)~=~-iG(\eta,\bta,J)~-~\int \dk 
~(\bta(k)\psi(k)+\bpsi(k)\eta(k)+J_{\mu 
}(-k)A^{\mu }(k)).
\eeq 
We recall the conjugated relations for the sources
\beq
\label {source}
\ \eta(k)~=~-(2\pi )^4 \frac {\delta \Gamma}{\delta \bpsi 
(k)}~~~~\bta(k)~=~(2\pi 
)^4 \frac 
{\delta 
\Gamma}{\delta \psi (k)}~~~~J_{\mu }(-k)~=~-(2\pi )^4 \frac {\delta 
\Gamma}{\delta A^{\mu 
}(k)}.
\eeq
From the invariance of the partition function (\ref{zp}) under infinitesimal 
translation of 
the 
fermion and gauge fields and under infinitesimal gauge transformation we deduce 
respectively the (DS) equations and the (WT) identities which are now 
mathematically 
well-defined objects.
The regularized form of the (DS) equations is
\beqa
\label {equamotA}
\ 0~&=&~ \bra \int\dk d\bar{p} d\bar{p'} ~\bpsi(p) \frac{\delta 
K(k,p,p')}{\delta 
A^{\nu }(r)} \psi(p')  
-~ \left( r^2 g_{\mu \nu } - r_{\mu } 
r_{\nu } (1-\frac{1}{\xi }) \right) \frac{A^{\mu }(-r)}{(2\pi )^4} \nonumber \\
& &+~\frac{J_{\nu }(-r)}{(2\pi )^4} ~+~ \frac{\delta S_{\Lambda }}{\delta 
A^{\nu 
}(r)} \ket \\
\label {equamotf}
\ 0~&=&~ \bra \int\dk d\bar{p'} ~K(k,p,p') \psi(p')~+~\eta (p) \ket,
\eeqa 
where the kernel $K(k,p,p')$ is explicitly and implicitly 
given 
by the expression of the action  (\ref{sktwo}) or (\ref{skthree}) respectively.
Due to the relations (\ref{source}) the regularized expression of the (WT) in 
terms 
of 
the generating functional of (1PI) functions can be written in momentum space 
as
\beq
\label {WTone}
(2\pi)^4r^{\mu}\frac {\delta \Gamma}{\delta A^{\mu }(r)}~=~e\rho_3(r) \int 
dk~ \left[ \bpsi(k) \frac {\delta \Gamma}{\delta \bpsi (k-r)}+\frac{\delta 
\Gamma}{\delta 
\psi (k)} \psi(k-r) \right] -\frac{1}{\xi} r^2r^{\mu} A_{\mu }(-r).
\eeq 
Except for the additional cutoff function $\rho_3$   the above 
expression 
is 
identical to the standard one.
Thus we expect that all general relations among Green's functions which can be 
deduced 
from 
(\ref{WTone}) are the same as in the case of known regularization schemes.
\mysection{The polarization operator and the 2n-photon amplitudes}
In many gauge invariant regularization schemes the calculation of 
amplitudes 
relative 
to closed fermion loops is a test of the method.
In fact any contribution to these amplitudes must be unambiguously finite.
We will see that this is indeed the case for the polarization operator and that 
due to 
explicit 
gauge invariance of the method this conclusion remains valid for the 2n-photon 
amplitudes. 

In order to calculate the polarization operator we take the functional 
derivative of the 
equation 
of motion (\ref{equamotA}) with respect to the C-number function $ A^{\mu }(-r' 
) $.
Then using (\ref{source}) to express $ J^{\nu }(-r) $, the inverse of the full 
photon 
propagator (\ref{photone}) is given 
in 
terms of the polarization operator $ \pi_{\mu \nu } $ by 
\beqa
\label {pione}
\ (2\pi )^4 \frac{\delta^2 \Gamma }{\delta A^{\mu }(-r' )\delta A^{\nu 
}(r)}~&=&~\pi_{\mu 
\nu }(r,r')~-~\delta (r-r')\left( r^2 g_{\mu \nu } - r_{\mu } r_{\nu } 
(1-\frac{1}{\xi }) \right)  \nonumber \\
\ & &+(2\pi )^4 \frac{\delta 
}{\delta 
A^{\mu }(-r' )} \bra \frac{\delta S_{\Lambda }}{\delta A^{\nu }(r)} \ket~ 
+~\mathcal{O}(A) 
\\ 
\label {pithree}
\ \pi_{\mu \nu }(r,r')~&=&~(2\pi )^4 \frac{\delta }{\delta A^{\mu }(-r' )}  
\int 
\dk 
d\bar{p} d\bar{p'} ~\bra\bpsi(p)\frac{\delta K(k,p,p')}{\delta 
A^{\nu }(r)} \psi(p') \ket.
\eeqa
Here $\mathcal{O}(A)$ represents terms which are vanishing when the sources are 
switched 
off.
We will now calculate the right hand-side of (\ref{pithree}) perturbatively at 
the one loop 
level.
The comparison of this result with the known expression of the regularized form 
of $ \pi_{\mu 
\nu 
} $ will then constrain the form of the operator $ S_{\Lambda } $.
Since in the action (\ref{skthree}) each monomial is proportional to the 
product 
of 
n 
gauge 
field\footnote{See the 
expression of $ F $ in appendix B. } and is of order $ e^n $, at order $ e^2$, 
the operator 
$\bpsi (\delta K / \delta A^{\nu } ) \psi $ is at most linear in the gauge 
field.
Hence using (\ref{skone}) and (\ref{skthree}) its vacuum 
expectation value can be 
expressed in 
terms of the full electron propagator 
\beq
\label {S}
\ S(p',p)~=~-i(2\pi )^4\frac{\delta^2 G}{\delta \bta (p') \delta \eta (p)},
\eeq
which is defined in presence of external sources, 
and only in terms of the vacuum expectation value of the gauge field.

Knowing that the functional derivative of $ S $ with respect to $ A^{\mu }(-r' 
) 
$ is at 
least of order $ e $, we obtain for $ \pi_{\mu \nu } $
\beqa
\label {pifour}
\ \pi_{\mu \nu }(r,r')~&=&~(2\pi )^4 Tr \Bigg 
\{ie \int  d\bar{p}~ 
\Gamma_{\nu }(p,p-r) \frac{\delta S(p-r,p)}{\delta A^{\mu }(-r')} \nonumber \\
\ & &~  -\frac{1}{2} e^2 \int \dk d\bar{p} d\bar{p'}~ \bigg [ \frac{r'_{\mu 
}}{r'^2}\rho_3(r') \left(\frac{\delta  
F_1}{\delta A^{\nu }(r)} (p-k+r',k-p') \right. \nonumber \\
\ & &~ -\left. \frac{\delta F_1}{\delta A^{\nu}(r)}(p-k,k-p'+r') \right) 
\rho_1^2(k)\kS~+~\rho_2(k) 
\rho_2(k+r') 
\rho_3(r') \Gamma_{\mu }(r')\nonumber \\
\ & &~ \frac{\delta 
F_1}{\delta A^{\nu }(r)}(p-k,k+r'-p') ~+~(r' \leftrightarrow -r,~\mu 
\leftrightarrow 
\nu ) \bigg ] S(p',p) \Bigg 
\} 
\eeqa
If $ \Gamma^{(2)} $ is the inverse of the full electron propagator, we can 
express $ 
\frac{\delta 
S }{\delta A^{\mu }} $ in terms of the three point function $ \frac{\delta 
\Gamma^{(2)} 
}{\delta A^{\mu }} $ (\ref{electhree}), and using (\ref{expanFpqO}) and 
(\ref{recur}) 
we 
get at order $ e $ 
\beqa
\label {pifive}
\ \frac{\delta S(p',p)}{\delta A^{\mu }(-r')}~&=&~\delta 
(p'-p+r')\frac{e}{(2\pi )^4}  
S(p')\Gamma_{\mu }(p',p) S(p).
\eeqa
Here $\Gamma_{\mu }(p',p)$ is defined in   (\ref{Gamone})  and $S(p)$ is the 
free 
electron propagator whose expression is readily deduced from the first term of 
(\ref{skthree}) and is given by 
\beq
\label {SO}
\ S(p)~=~  \frac{1}{\rho_1^2 (p) \pS-m}.
\eeq
Due to the properties (\ref{Gamtwo}) and (\ref{recur2}) we can notice that the 
expression 
(\ref{pifour}) converges formally to the non 
regularized 
form of 
the polarization operator when the cutoff $ \Lambda $ goes to infinity.
 
Now if we substitute in (\ref{pifour}) $  \delta S / \delta A^{\mu } $ by its 
expression 
(\ref{pifive}), we can see that the polarization operator
\beq
\label {piix}
\pi_{\mu \nu }(r)\delta (r-r')~\equiv ~\pi_{\mu \nu }(r,r') 
\eeq
has the following structure
\beq
\label {structone}
\ \pi_{\mu \nu }(r)~=~Tr \left[ r_{\mu } r_{\nu } A~+~\left(r_{\mu} \Gamma_{\nu 
}(r)B_1+r_{\nu} \Gamma_{\mu 
}(r)B_2 \right)~+~\Gamma_{\mu }(r) \otimes \Gamma_{\nu }(r)C \right].
\eeq
Since the theory is gauge invariant by construction, the (WT) 
(\ref{WTone}) 
tells 
us 
that the polarization operator must be transverse.
Owing to the transversality property (\ref{cont}) of the $ \Gamma_{\mu } $ 
matrices, 
this implies that the matrices $A$, $B_1$ and $B_2 $ of (\ref{structone}) must 
be 
zero 
if we multiply both 
sides of 
(\ref{structone}) successively by $ r ^{\mu} r^{\nu} $, and then by $ r^{\mu} $ 
and 
$ r^{\nu} $.
This is indeed the case.
A straightforward calculation shows that $B_1$ and $B_2 $ are both proportional 
to 
the 
integral
\beq
\label {Bint}
\int dp ~ \rho_2(p)\left( \rho_2(p+r)S(p)- \rho_2(p-r)S(p-r) \right),
\eeq
and that the matrix $A$ is proportional to the integral
\beq
\label {Aint}
\int dp ~ \left[ \left( \rho_1^2(p)\pS-m \right) S(p-r)-\left( \rho_1^2(p+r) 
(\pS +\rS)-m \right)S(p) \right].
\eeq
Since for high squared momenta the asymptotic form of the free electron 
propagator 
(\ref{SO}) is $1/m$ the two integrals (\ref{Bint}) and (\ref{Aint}) are 
regularized 
and vanish identically after a shift of variable.
However there is a great difference between the matrix  $A$  and the matrices 
$B_1$ 
and $B_2$.
Whereas in the massless case one can  always  choose the cutoff functions in 
order 
to 
regularize the matrices $B_1$ and $B_2$ given by (\ref{Bint}), this is not true 
for  
$A$ where now terms like $\rho_1^2(p) \rho_1^{-2}(p-r)$ appear in the integral 
(\ref{Aint}).
In contradistinction to the massive case, it seems very difficult to construct 
a 
continuous 
non-perturbative regularization scheme for massless fermions which is also 
mathematically clean.
This is reminiscent of the Nielsen-Ninomiya Theorem \cite{NNT} which implies 
that 
a 
chiral invariant regularization is still lacking in the fermionic sector of 
lattice gauge theory.

\begin{figure}
	\epsfxsize=8cm
	\epsfysize=3.5cm
	\centerline{\epsffile{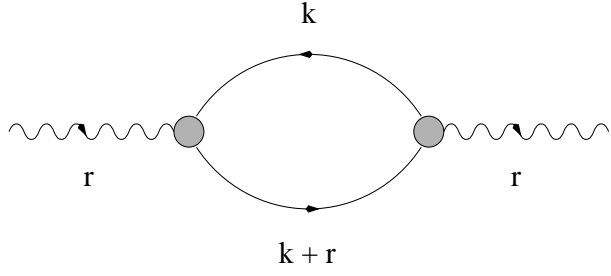}}
\caption{The polarization operator.\label{polar1}}
\end{figure}

We thus obtain for the regularized form of the polarization operator 
(\ref{pithree})
\beq
\label {pisix}
\pi_{\mu \nu }(r)~=~i \frac{e^2}{(2\pi )^4} \rho_3^2(r)  Tr 
\int dk~ \bigg[\rho_2^2 (k)\rho_2^2 (k+r)
 \Gamma_{\mu }(r) S(k+r) \Gamma_{\nu }(r) S(k)\bigg].
\eeq
The polarization operator  can be represented by the diagram Fig.\ref{polar1} where all 
the vertices are transverse Fig.\ref{vert1}b.
In the present work we calculate the relevant (1PI) functions with cutoff 
functions 
of the form
\beq
\label {rhok}
\ \rho(k)~=~e^{ \frac{k^2}{\Lambda^2} }, 
\eeq
$ \Lambda $ being the (UV) cutoff.
In order to  simplify the calculation we choose,
\beq
\label {relcut}
\rho_1^4(k)~=~\rho_2(k),~~~~~~\rho_2(k)~=~\rho_3(k)~=~\rho(k).
\eeq
In this case we can easily show that
\beq
\label {intmajor}
\int dk~f(k,\Lambda) S(k)~=~\int dk~f(k,\Lambda) \rho^{-1}(k) \frac 
{\sqrt{\rho(k)} 
\kS +m}{k^2(1+\frac{m^2}{\Lambda^2})-m^2}~+~\mathcal{O}( \frac {1}{\Lambda} )
\eeq
provided that $|f(k,\Lambda)|$  behaves like $ \rho^2(k) /k$ for high $k^2$.
This implies that we can always make the replacement
\beq
\label {repSO}
S(k) \rightarrow (1-\frac {m^2}{\Lambda^2}) \rho^{-1}(k)\frac {\sqrt{\rho(k)} 
\kS 
+m}{k^2-m^2}
\eeq
in the integrals which  we encounter in the calculation of the relevant (1PI) 
functions without changing the result of the integration.

If we take the trace after the substitution (\ref{repSO}) and use the Feynman 
parametrization for the product of 
propagators the 
expression (\ref{pisix}) can be written as
\beqa
\label {piseven}
\ \pi_{\mu \nu }(r)~&=&~4i\rho^2(r) \frac{e^2}{(2\pi )^4} \frac{1}{r^4} 
(1-2\frac{m^2}{\Lambda^2}) 
\int_{0}^{1} dx \int 
dk~\frac{1 }{ 
[-k^2-2krx+m^2-r^2x]^2 
} \Bigg \{ e^{\frac{3}{2} \frac{k^2}{\Lambda^2} } e^{\frac{3}{2} 
\frac{(k+r)^2}{\Lambda^2} 
} \nonumber \\
\ && 2\bigg[(kr)^2r_{\mu }r_{\nu }-r^2kr(r_{\mu }k_{\nu }+r_{\nu }k_{\mu 
}) 
+r^4k_{\mu }k_{\nu } - \frac{r^2}{2} (r^2g_{\mu \nu }-r_{\mu }r_{\nu 
}) (k^2+kr)\bigg] \nonumber \\
\ && + e^{ \frac{k^2}{\Lambda^2} } e^{ \frac{(k+r)^2}{\Lambda^2} } 
m^2r^2(r^2g_{\mu 
\nu 
}-r_{\mu }r_{\nu 
})  \Bigg \}
\eeqa
The asymptotic form of the integrals which contribute to the right hand-side 
of (\ref{piseven}) are given in appendix D.
With the results (\ref{intone}) and (\ref{inttwo}) and if we introduce an 
arbitrary 
unit 
of 
mass $ \mu $, (\ref{piseven}) can be rewritten 
\beqa
\label {pieight}
\ \pi_{\mu \nu }(r)~&=&~-\frac{\alpha}{\pi} (r^2 g_{\mu \nu } - 
r_{\mu } r_{\nu 
}) \bigg\{ 
\frac{1}{2r^2} \left[  \frac{\Lambda^2}{3}-m^2\left( \frac{5}{3} 
-2\log{\frac{3}{2}} 
\right)
\right] + \frac{2}{3} \log{\frac{\Lambda}{\mu} } -\frac{\gamma}{3}+ 
\frac{29}{72}-\frac{1}{3} \log 3 \nonumber \\
\ &&  -2 \int_{0}^{1} dx~x(1-x)\log{ \frac{ m^2-r^2x ( 1-x )  }{\mu^2} } \bigg 
\} , 
\eeqa
$ \alpha $ being the fine structure constant.
We see from the expression (\ref{pieight}) that the polarization operator is 
transverse 
as expected and has the standard form \cite{PO}, except for the quadratic 
divergent 
term 
which is 
proportional to $ 1/r^2 $.
As noticed in Section 2 the origin of this term is due to the fact that in 
configuration 
space the regularized action $S_{Reg}-S_{\Lambda}$ (\ref{actferm}) and 
(\ref{actgau}) 
converges 
exactly to the non-regularized action of QED if one makes the assumption that 
taking 
the 
limit commutes with integration. 
Stated differently, this discrepancy is due to the non-uniform convergence of 
the 
integral 
in 
momentum space.
The new divergent term of (\ref{pieight}) can be 
removed by adding the counterterm
\beq
\label {Slamb}
\ S_{\Lambda}~=~\frac{\alpha}{4\pi} \int \dk ~\rho_3^2(k) \frac{1}{k^2} ( k^2 
g_{\mu 
\nu } - k_{\mu } k_{\nu } ) \left[  \frac{\Lambda^2}{3}-m^2\left( \frac{5}{3} 
-2\log{\frac{3}{2}} 
\right)
\right]   A^{\mu }(k)A^{\nu }(-k) 
\eeq
to the expression of the regularized action $S_{Reg}$ in momentum space.
The counterterm $ S_{\Lambda} $ can be viewed as associated to a  new bare 
interaction
\beq
\label {Sc}
\ S_{C}~=~c\frac{\alpha}{4\pi} \int \dk ~\rho_3^2(k) \frac{1}{k^2} ( k^2 g_{\mu 
\nu } - k_{\mu } k_{\nu } )  A^{\mu }(k)A^{\nu }(-k), 
\eeq
where the coupling constant $c$ has the dimension of the square of a mass.
This interaction which is non-local in configuration space can be absorbed in 
the 
inverse of the photon 
propagator (\ref{pione}).
Since at order $\alpha^n$ the photon propagator (\ref{photone}) will now 
contain a term 
proportional to $c^n/(k^2)^{(n+1)}$, the contributions of this interaction to 
the 
potentially 
divergent (1PI) functions are all (UV) finite.
This implies that the new coupling constant $c $, which can give masses to the 
photon, is not renormalized by higher order 
corrections.
It follows that the counterterm $ S_{\Lambda} $ is uniquely defined by the 
choice of the renormalized form of the coupling constant $c $.
In order to keep the photon massless as in standard QED and to avoid any 
non-local 
interaction in the action, we restrict ourselves 
to 
the form (\ref{Slamb}) of the counterterm, which means 
that the renormalized form of the coupling constant $c $ is tuned to zero.
This gives the known result for the polarization operator (\ref{pithree}).
Another choice for the bare coupling constant $c$, that is to say for the 
counterterm $ 
S_{\Lambda} $, seems to open  the possibility of  dynamical mass generation   
for the 
photon through the  non-local interaction (\ref{Sc}). 
 
Now we show that the expressions for the 2n-photon amplitudes are standard.
We begin by the calculation of the polarization tensor.
If we use the equation of motion (\ref{equamotA}) and the definitions 
(\ref{source}), 
since $ 
S_{\Lambda } $ is quadratic in the gauge field, the (1PI) function relative to 
the four 
photons 
amplitudes is given by
\beqa
\label {tenone}
\ \frac{\delta^4 \Gamma}{\delta A^{\beta } (r_4) \delta A^{\alpha } (r_3) 
\delta A^{\nu 
} 
(r_2) 
\delta A^{\mu } (r_1)}~&=& -i(2\pi)^4  \frac{\delta^3}{\delta 
A^{\beta } (r_4) \delta A^{\alpha } (r_3) \delta A^{\nu } (r_2) }~ Tr \int \dk 
d\bar{p} 
d\bar{p'}~\frac{\delta 
K}{\delta A^{\mu }(r_1)}(k, p, p',\frac{\delta }{\delta J}) \nonumber \\
\ & & S(p',p)e^G~+~\mathcal{O}(A), 
\eeqa
where the differential operator $ \delta K/ \delta A^{\mu } $ with respect to 
the gauge 
field 
sources $ J $ is easily obtained from (\ref{skone}), (\ref{skthree}), 
(\ref{expanFpq}) and 
(\ref{expanFpqO}). 

\begin{figure}
	\epsfxsize=7cm
	\epsfysize=5cm
	\centerline{\epsffile{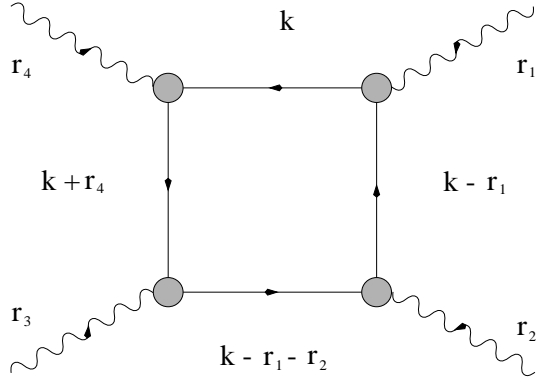}}
\caption{The representation of the amplitude $A_{\mu \nu \alpha \beta } (r_1, r_2, r_3, 
r_4)$.\label{box1}}
\end{figure}

In the last expression it is understood that $ \delta K/ \delta A $ acts  
first on 
the 
product $ Se^G $ and then that we act successively with the functionals 
derivatives 
$ 
\delta / \delta A $.
Knowing that the action of $ \delta / \delta J $ or $ \delta / \delta A $ on $ 
S $ 
(\ref{S}) 
gives rise to terms which are at least of order $ e $\footnote{See appendix C. 
} and 
that 
at 
the one loop order terms of the order $ \mathcal{O}(e^5) $ are neglected the 
polarization tensor
\beq
\label {tentwo}
\ \Gamma^{(4)}_{\mu \nu \alpha \beta } (r_1, r_2, r_3, 
r_4)\delta(r_1+r_2+r_3+r_4)~\equiv 
~((2\pi)^4)^3 
\frac{\delta^4 \Gamma}{\delta A^{\beta } (r_4) \delta A^{\alpha } (r_3) \delta 
A^{\nu } 
(r_2) 
\delta A^{\mu } (r_1)} 
\eeq
can be written as,
\beqa
\label {tenthree}
\ \Gamma^{(4)}_{\mu \nu \alpha \beta } (r_1, r_2, r_3, r_4)~&=&~Tr \bigg[ 
r_{\mu 1 } 
r_{\nu 
2} 
r_{\alpha 3} r_{\beta 4}A~+~(r_{\nu 2} r_{\alpha 3} r_{\beta 4} \Gamma_{\mu 
}(r_1)B_1+three~similar)\nonumber \\
\ & &+~(r_{\alpha 3} r_{\beta 4}\Gamma_{\mu }(r_1) \otimes \Gamma_{\nu 
}(r_2)C_1+five~ 
similar) 
\nonumber \\
\ & &+~(r_{\beta 4} \Gamma_{\mu }(r_1) \otimes \Gamma_{\nu }(r_2) \otimes 
\Gamma_{\alpha 
}(r_3) 
D_1+three ~similar)\nonumber \\
\ & &+~\Gamma_{\mu }(r_1) \otimes \Gamma_{\nu }(r_2) \otimes \Gamma_{\alpha 
}(r_3) 
\otimes
\Gamma_{\beta }(r_4)E \bigg].
\eeqa
Here $A$, $B_i$, $C_i$, $D_i$ and $E$ are scalar functions of the four momenta 
$r_1$, 
$r_2$, 
$r_3$ and $r_4$ and tensorial expressions with respect to the spin indices.
Since the regularized polarization tensor is gauge invariant by construction, 
it 
follows 
that $A$ is identical to zero if both sides of (\ref{tenthree}) are multiplied 
  by the product $ r_1^{\mu } r_2^{\nu } 
r_3^{\alpha} 
r_4^{\beta}$.
Similarly we can show that only the scalar function $E$ in (\ref{tenthree}) is 
non 
vanishing. 
The structure of $ \Gamma^{(4)}_{\mu \nu \alpha \beta } $ being now 
established, we 
keep 
only 
the term proportional to $ \Gamma_{\mu } $ in $ \delta K/ \delta A $.
Since this term is not a differential operator the calculation of the right 
member of 
(\ref{tentwo}) is indeed reduced to the evaluation of $\delta / \delta A^{\beta 
} 
\delta 
/ 
\delta A^{\alpha } \delta / \delta A^{\nu }~ S $.
If we define 
\beqa
\label {tenfour}
\ A_{\mu \nu \alpha \beta } (r_1, r_2, r_3, r_4)&\equiv&i\frac{e^4}{(2\pi 
)^4}\rho_3(r_1) \rho_3(r_2)\rho_3(r_3)\rho_3(r_4)  Tr \int 
dk~\rho_2^2(k)\rho_2^2(k-r_1)\nonumber \\
\ & & \rho_2^2(k-r_1-r_2) \rho_2^2(k+r_4) \bigg[ \Gamma_{\mu 
}(r_1) 
S(k-r_1) 
 \Gamma_{\nu }(r_2) S(k-r_1-r_2)\nonumber \\
\ & & \Gamma_{\alpha }(r_3)S(k+r_4)  \Gamma_{\beta }(r_4) S(k) \bigg]
\eeqa
and use the result (\ref{tenthree}) and the definitions (\ref{relcut}) for the 
cutoff 
functions we 
find for the polarization tensor Fig.\ref{box1}
\newpage
\beqa
\label {tenfive}
\ \Gamma^{(4)}_{\mu \nu \alpha \beta } (r_1, r_2, r_3, 
r_4)~&=~&  A_{\mu \nu \alpha 
\beta } 
(r_1, 
r_2, r_3, r_4) +A_{\mu \nu \beta \alpha } (r_1, r_2, r_4,r_3) +A_{\mu \beta \nu \alpha 
} 
(r_1, 
r_4, 
r_2, 
r_3)\nonumber \\
\ & & +A_{\mu \alpha \nu \beta } (r_1, r_3, r_2, r_4)+A_{\mu \alpha \beta \nu } (r_1, 
r_3, r_4, r_2)\nonumber \\
\ & & +A_{\mu \beta \alpha \nu } 
(r_1, 
r_4, 
r_3, 
r_2) .
\eeqa
Now we are ready to show that $ \Gamma^{(4)}_{\mu \nu \alpha \beta } $ is as 
expected 
finite 
\cite{KA} and is regular in the (IR) domain.
After the substitution (\ref{repSO}) and a judicious shift of the integration 
variable in 
(\ref{tenfour}), the 
product of 
the 
cutoff functions becomes proportional to $ \rho^6(k) $, 
where 
 $ \rho(k) $ is defined in (\ref{rhok}).
From the rotational invariance of the integral and the properties (\ref{comm}) 
and 
(\ref{cont}) 
 of the $\Gamma_{\mu }$ matrices, it follows that the asymptotic expression of 
(\ref{tenfour}) 
as $-q^2 \rightarrow +\infty$ is 
\beq
\label {tensix}
\ A_{\mu \nu \alpha \beta }~\sim ~Tr~(2\Gamma_{\mu} \Gamma_{\nu} 
\Gamma_{\alpha} 
\Gamma_{\beta}-\Gamma_{\alpha} 
\Gamma_{\mu}\Gamma_{\nu}\Gamma_{\beta}-\Gamma_{\beta} 
\Gamma_{\mu}\Gamma_{\alpha}\Gamma_{\nu}) \int dk~ \rho^6(k) 
\frac{k^4}{(m^2-k^2)^4}.
\eeq
As a corollary it is readily seen that the potentially logarithmic divergent 
term of 
(\ref{tenfive}) vanishes identically in taking the sum, i.e. the polarization 
tensor is 
finite.
Is the polarization tensor regular?

The standard one loop expression of the polarization tensor $\Gamma^{(4)}_{S 
\mu \nu 
\alpha 
\beta }$ 
which 
is known to be finite and gauge invariant \cite{KA}, can be deduced from 
(\ref{tenfour}) 
and 
(\ref{tenfive}) if one substitutes respectively the matrices $ \Gamma_{\mu}$ by 
the 
Dirac's 
matrices $\gamma_{\mu}$. 
As a result, the structure of the difference $ \Gamma^{(4)}_{ \mu \nu \alpha 
\beta 
}- 
\Gamma^{(4)}_{S \mu \nu \alpha \beta } $ is given by (\ref{tenthree}) after the 
substitution 
of the 
matrices $ \Gamma_{\mu}$ by Dirac's matrices 
$\gamma_{\mu}$, if in this expression the 
last 
term in the right member is omitted. 
Since this difference is actually finite and gauge invariant, we can show along 
the 
same 
lines as before that it vanishes identically.
Therefore the polarization tensor has the standard form, and thus is regular. 
Since for $ n > 2$ the 2n-photon amplitudes are finite, we can show in a 
similar way 
that 
they are 
also regular.
We end this section by discussing the status of the n-photon amplitudes when n 
is odd.

If $S$ is the regularized electron propagator (\ref{SO}) and if the matrix 
$\Gamma_{\mu}$ 
is given by (\ref{gam}), the requirement of gauge invariance imposes that the 
n-photon 
amplitude, whose structure is similar to (\ref{tenthree}), must be proportional 
to 
the 
integral of the trace of $n$ products of 
$\Gamma_{\mu}S$.
Under charge conjugation the matrices $\Gamma_{\mu}$ transform exactly in the 
same way 
as Dirac's matrices $\gamma_{\mu}$.
Hence it follows by virtue of Furry's theorem that the n-photon amplitude 
vanishes when 
$n$ 
is odd.
\mysection{The mass operator and the vertex function}
First we derive the expression of the mass operator at the one loop level.

For this sake we take the functional derivative of the equation of motion 
(\ref{equamotf}) defined by (\ref{skone}) and  (\ref{skthree}) 
with respect to the C-number function $ \psi^b(p') $ and express 
the 
sources 
of 
the 
electron field according to the generating functional of (1PI) functions by 
relations 
(\ref{source}).
We thus obtain the inverse of the full electron propagator $ \Gamma^{(2)} $ 
(\ref{elecone}) 
in 
terms 
of 
the 
mass operator $ \Sigma $ as
\beqa
\label {massone}
\ \Gamma^{(2)}_{ab}(p,p')~&=&~(\rho_1^2(p) \pS-m)_{ab} \delta 
(p-p')-\Sigma_{ab} (p,p') 
\\
\label {masstwo}
\ \Sigma_{ab} (p,p')~&=&~-ie\int dk 
dk'\Gamma^{(2)}_{db}(k',p')\Gamma^{\mu}_{ac}(p,k)\frac{\delta}{\delta J^\mu 
(k-p)}S_{cd}(k,k')e^G \nonumber \\
\ & & - \sum_{n=0}^{+\infty} \frac{(ie)^{n+2}}{(n+2)!} \int \dk 
d\bar{k'}dp''~\Gamma^{(2)}_{db}(p'',p') 
\bigg[ F_{n+2}(p-k,k-k')\rho_1^2(k) \kS
\nonumber 
\\
\ & &+(n+2)\int dq~\rho_2(k) \rho_2(k-q)\rho_3(q) \Gamma_{\mu}(q) 
\frac{\delta}{\delta 
J_\mu (-q)} F_{n+1}(p-k,k-k'-q) \bigg]_{ac}\nonumber \\
\ & & S_{cd}(k',p'')e^G~+~\mathcal{O}(\bpsi) .
\eeqa
Here the  $ F_n(p,q) $ which are  defined in appendix B are expressed in terms 
of the 
functional derivative $ \delta /\delta J$ and the latin letters refer to the spin 
indices.
The contribution of order $e^2$ which 
comes from 
the 
bracket of expression (\ref{masstwo}) is proportional to $ 
\Gamma_{\mu}(q)q_{\nu}D^{\mu\nu}(q) $ 
and  
vanishes owing to the properties (\ref{cont}).
Taking into account the fact that the functional derivative $ \delta S/\delta 
J_{\mu} $ 
can be 
written 
in 
terms of the product of the full photon propagator 
\beq
\label {D}
\ D^{ \mu \nu }(r,q)~=~i(2\pi)^4\frac{\delta^2 G}{\delta J_\mu (r) \delta J_\nu 
(-q )}
\eeq
times the functional derivative $ \delta S/\delta A^{\nu} $ whose expression at 
order 
$e$ 
is 
given 
by (\ref{pifive}), we get for the mass operator at order $e^2$  
\beq
\label {massthree}
\ \Sigma(p,p')~=~\delta (p-p')i\frac{e^2}{(2\pi)^4} \int dk~D^{ \mu \nu 
}(k-p)\Gamma_{\mu}(p,k)S(k) 
\Gamma_{\nu}(k,p).
\eeq
\begin{figure}
	\epsfxsize=7cm
	\epsfysize=3cm
	\centerline{\epsffile{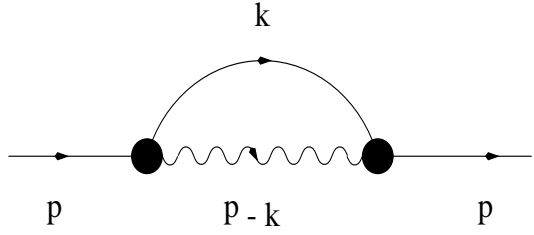}}
\caption{The mass operator.\label{mass1}}
\end{figure}
This operator can be represented with the help of the vertex Fig.\ref{vert1}a by the 
diagram of Fig.\ref{mass1}.
In this expression $D^{\mu\nu}(k)$ is the free photon propagator whose 
expression
\beq
\label {photfree}
\ D_{\mu\nu}(k)~=~-  \frac{1}{k^2} \left( 
g_{\mu\nu}-\frac{1}{k^2} 
k_{\mu}k_{\nu}(1-\frac{1}{\xi}) \right)
\eeq
is deduced from (\ref{sgauge}) and $\Gamma_{\mu}(p,k)$ is 
defined in 
(\ref{Gamone}).
Owing to the property (\ref{Gamtwo}) of the matrix $\Gamma_{\mu}(p,k)$, we can 
notice 
that 
we recover the standard non regularized form for the mass 
operator as 
the 
cutoff 
$\Lambda$ tends to infinity.
The mass operator $\Sigma $ can be written as
\beqa
\label {masfour}
\ & & \Sigma(p) \delta (p-p')~ \equiv ~\Sigma (p-p') \\
\label {massfour}
\  & & \Sigma(p)~=~A(p^2)+\pS B(p^2). 
\eeqa
As stated before, we can replace freely the free propagator $S(k)$ by the 
expression 
(\ref{repSO}) without changing the result of integration in (\ref{massthree}).
Then if we use definitions (\ref{relcut}) and (\ref{rhok}) for the cutoff 
functions 
and 
neglect finite 
terms 
of 
the 
order $ \mathcal{O} (1/\Lambda )$ we get for the scalar functions $A$ and $B$
\beqa
\label {massfive}
\ A(p^2)&=&-im\frac{e^2}{(2\pi)^4} \int dk~ \bigg \{ \rho^2(k-p)
\frac{1}{\left[ 
\kappa^2-(k-p)^2 \right]^2}~+~3\rho^2(k-p)\rho(k) \frac{1}{\left[ 
\kappa^2-(k-p)^2 
\right] 
(m^2-k^2) 
}\nonumber \\ 
\ & &-~\frac{p^2-2kp+m^2}{\left[ \kappa^2-(k-p)^2 \right ]^2 
(m^2-k^2) } 
\bigg 
\} 
\\
\label {masssix}
\ p^2B(p^2)&=&i\frac{e^2}{(2\pi)^4} \int dk~ \rho^2(k-p)\left \{ 
2\rho^{\frac{3}{2}}(k) 
\frac{kp}{ 
\left[ 
\kappa^2-(k-p)^2 \right ] (m^2-k^2) } ~+~ \left[ kp\left(\rho^{\frac{3}{2}}(k)- 
\rho^{\frac{1}{2}}(k)\right) \right.\right. \nonumber \\ 
\ & &   +\left. \left. 2p^2\left(1-\rho^{\frac{3}{2}}(k)\right)\right] 
\frac{1}{ 
\left[ \kappa^2-(k-p)^2 \right ]^2   } \right \}
\eeqa
where the last integral of the right member of (\ref{massfive}) is finite.
In order to avoid the problem of potentially (IR) divergences due to the 
momentum 
carried 
by 
the 
internal photon line, we give an arbitrary small mass $\kappa $ to the photon. 
After Feynman's parametrization of the product of propagators, with the help of 
formulas 
(\ref{intzero}), (\ref{intone}) and  (\ref{intthree}), the integration over the 
four 
momentum is 
easily performed and we obtain in the Feynman gauge the known results \cite{PO}
\beqa
\label {massseven}
\  A(p^2)&=&m\frac{\alpha}{\pi}\left  \{2\log{\frac{\Lambda }{\mu }}-\gamma 
-1-\frac{1}{4}\log 
54- \int_{0}^{1} dx~\log{ \frac{-p^2x(1-x)+m^2(1-x)+x\kappa^2 }{\mu^2} }\right 
\} 
\\
\label {masseight}
\ B(p^2)&=&-\frac{\alpha}{2\pi}\bigg \{ \log{\frac{\Lambda }{\mu }} 
-\frac{\gamma }{2}- 
\frac{73 }{140} - \frac{1 }{2} \log {\frac{8}{5}} \nonumber \\
\ & &-\int_{0}^{1} dx~x\log{  \frac{-p^2x(1-x)+m^2(1-x)+x\kappa^2 }{\mu^2} 
}\bigg  \}.
\eeqa

\begin{figure}
	\epsfxsize=5cm
	\epsfysize=6cm
	\centerline{\epsffile{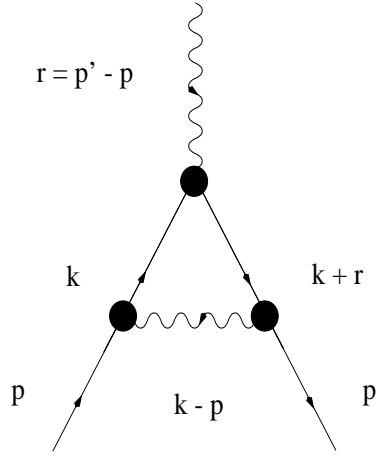}}
\caption{The vertex function.\label{vert2}}
\end{figure}

We now calculate the vertex function at the one loop level.
If we define the (1PI) function $\Gamma^{(3)}$ as,
\beq
\label {verone}
\  ((2\pi)^4)^2 \frac{\delta^3 \Gamma }{\delta A^{\mu }(-r )\delta \bpsi (p) 
\delta 
\psi 
(p')}~=~  \Gamma^{(3)}_{\mu}(p,p',r)~\equiv ~e\delta 
(p'-p-r)\Gamma^{(3)}_{\mu}(p,p')
\eeq
in the same way as we deduced the mass operator, we get from the equation of 
motion 
(\ref{equamotf}), (\ref{skone}) and (\ref{skthree})
\beq
\label {vertwo}
\ \Gamma^{(3)}_{\mu}(p,p',r)~=~\delta (p'-p-r)e\Gamma_{\mu}(p,p')~+~ 
(2\pi)^4\frac{\delta}{\delta A^{\mu }(-r ) } \left( \Sigma (p,p') +\Sigma ' 
(p,p') 
\right).
\eeq
In this expression 
\beq
\label {verthree}
\ \Sigma(p,p')~=~ i\frac{e^2}{(2\pi)^4} \int dkdq~D^{ \alpha\beta } 
(k-p,q)\Gamma_{\alpha 
}(p,k)S(k,p'+q) \Gamma_{\beta  }(p'+q,p')
\eeq
is the mass operator at the one loop order, $\Gamma_{\mu}(p,p')$ is given by 
the 
definition (\ref{Gamone}) and
\beqa
\label {verthreebis}
\ \Sigma ' (p,p')&=& \int \dk d\bar{k'}dp''~\Gamma^{(2)}_{db}(p'',p') \Bigg\{ 
e^2\bigg[  
\frac{1}{2} F_2(p-k,k-k')\rho_1^2(k) \kS \nonumber \\ 
\ & & +\int dq~\rho_2(k) \rho_2(k-q)\rho_3(q) \Gamma_{\nu}(q)  F_1(p-k,k-k'-q) 
\frac{\delta}{\delta J_\nu (-q)} \bigg]_{ac} S_{cd}(k',p'')\nonumber \\
\ & &+~ e^3\frac{i}{2} \bigg[ \frac{1}{3} F_3(p-k,k-k')\rho_1^2(k)  \kS
+\int 
dq~\rho_2(k) \rho_2(k-q)\rho_3(q) \Gamma_{\nu}(q)\nonumber \\ 
\ & & F_2(p-k,k-k'-q) \frac{\delta}{\delta J_\nu (-q)} \bigg]_{ac} 
S_{cd}(k',p'')+\mathcal{O}(e^4) \Bigg\}e^G.   
\eeqa
If we keep only the terms of order $e^3$ in the expression (\ref{verthreebis}) 
and 
use the recursion formulas (\ref{recur}) we can show respectively that the 
first and 
the 
second bracket of $\Sigma ' $ vanish on the one hand 
after 
integration over $dk$ and on the other hand after successive  integration over 
$dp''$,$dk'$ 
and 
$dk$.
Then according to the relation (\ref{pifive}), the functional derivative of  $ 
\Sigma 
$ (\ref{verthree}) with respect to the C-number function $ A^{\mu} $ is easily 
obtained.
We thus get for the vertex function Fig.\ref{vert2} at the one loop level
\beq
\label {verfour}
\ \Gamma^{(3)}_{\mu}(p,p')~=~ \gamma_{\mu}~+~\frac{ie^2}{(2\pi)^4} \int 
dk~D^{\alpha 
\beta}(k-p) \Gamma_{\alpha }(p,k) S(k) 
\Gamma_{\mu}(k,k+r)S(k+r)\Gamma_{\beta}(k+r,p').
\eeq

As the (UV) cutoff $\Lambda$ tends to infinity due to the property 
(\ref{Gamtwo}), we 
recover formally the standard non regularized expression for the vertex 
function.

Now we work in Feynman gauge and assume that the external electron lines are on 
mass-shell, thus we use freely the Gordon relation and express as usual the 
vertex 
function as
\beq
\label {verfive}
\ \Gamma^{(3)}_{\mu}(p,p')~=~\gamma_{\mu} \left( 1+F_1(r^2) \right) 
~+~\frac{i}{2m} 
\sigma^{\mu \nu} r_{\nu}F_2(r^2),
\eeq
where $r=p'-p$ is the momentum transfer. 
In this case the computation can be simplified if one notices that the term $ 
\rho_1^2(p) 
\pS-\rho_1^2(k) \kS$ which occurs in $\Gamma_{\mu }(p,k)$ (\ref{Gamone}) is 
identical 
to 
$S^{-1}(p)-S^{-1}(k)$ and that
\beq
\label {Gamasymp}
\ \bu (p) S^{-1}(p) ~=~\bu (p) \mathcal{O} 
(\frac{1}{\Lambda^2}),
\eeq
$p$ being the momentum of the outgoing electron.
A similar reduction occurs for the term $\Gamma_{\beta}(k+r,p')u(p')$.
We give to the photon an arbitrary small mass $\kappa$ in order to avoid the 
(IR) 
divergence and substitute the free electron propagators by their expressions 
(\ref{repSO}).
After Feynman parametrization of the product of propagators, we 
perform  
the integration over the internal momentum using the formulas of appendix C.
Then, if we define the integrals 
\beqa
\label {defint}
\ I_1(r^2)&=&\int_{0}^{1} 
\frac{dx}{m^2-r^2x(1-x)}~~~~~~I_3(r^2)~=~\int_{0}^{1}~dx 
\log{ 
\frac{m^2-r^2x(1-x)}{\mu^2}}\nonumber \\
\ I_2(r^2)&=&\int_{0}^{1} \frac{dx}{m^2-r^2x(1-x)} \log{ 
\frac{m^2-r^2x(1-x)}{\mu^2}}
\eeqa
we get for the forms factors $F_1(r^2)$ and $F_2(r^2)$ the expressions
\beqa
\label {versix}
\ F_1(r^2)&=&\frac{\alpha}{\pi} \Bigg\{ \frac{1}{2}\log{\frac{\Lambda 
}{\mu}}-\frac{1}{4} 
\gamma -\frac{3}{8}-\frac{1}{2}\log 2-\frac{1}{4}(2m^2-r^2)I_2(r^2)+\frac{1}{2} 
\bigg[ 
(2m^2-r^2) 
\log{\frac{\kappa }{\mu}}\nonumber \\
\ & &+ (3m^2-r^2)\bigg]I_1(r^2)-\frac{1}{4}I_3(r^2) \Bigg\} \\
\label {verseven}
\ F_2(r^2)&=&\frac{\alpha}{2\pi}m^2I_1(r^2).
\eeqa
$\mu $ being an arbitrary unit of mass.
As expected from (\ref{verfive}) and (\ref{verseven}) we recover the known 
result for 
the 
anomalous magnetic moment of the electron \cite{PO}.
The expression of the form factor $F_1(r^2)$ which is (IR) and (UV) divergent 
is also 
standard.

What about the prediction of the (WT) identity?
From the general relation (\ref{WTone}) we can deduce the identity
\beq
\label {WTtwo}
\ 
r^{\mu}\Gamma^{(3)}_{\mu}(p,p',r)~=~e\rho_3(r) 
\left[\Gamma^{(2)}(p+r,p')-\Gamma^{(2)}(p,p'-r) \right],~~~~r=p'-p.
\eeq
If $Z_1$ and $Z_2$ are respectively the renormalization constants of the vertex 
function 
and of the external electron lines associated with the mass operator i.e.
\beq
\label {renone}
Z_2-1~=~- \left. \frac{\partial \Sigma(p)}{\partial \pS} \right |_{\pS=m}
\eeq
the identity (\ref{WTtwo}) implies that $Z_1Z_2^{-1}$ must be finite.
If the external electron lines are off mass-shell, we can show that a term 
proportional 
to   
$ \alpha \rS r_{\mu}/r^2$ contributes to the vertex function too.
Since the derivative of this term is not regular near the origin, the 
expression $r^{\mu}\partial\Gamma^{(3)}_{\mu}/\partial r^{\rho} $ does not 
vanish as 
$r^{\mu} \rightarrow 0$.
As a consequence, the (WT) (\ref{WTtwo}) does not imply the equality of $Z_1$ 
and $Z_2 
$.
This explains the reason why $Z_1$ differs from $Z_2$ in first order in 
$\alpha$ by a 
numerical 
constant 
as it 
can be easily seen.
\mysection{Conclusion}
We have constructed in four dimensions   a  continuous regularization scheme of 
QED based on a 
non 
local 
extension of the action.
Since the measure of the path integral is invariant under the gauge group, this 
regularization scheme is actually non-perturbative.
Once the regularized action is fixed by the choice of the cutoff functions, the 
regularized amplitudes can be calculated in a straightforward way from the path 
integral 
without the need to adjust some parameters or to define some integrals 
formally.
In order to illustrate this fact we have deduced the regularized (1PI) 
functions at the one loop order from the equations of motion, 
which are now mathematically  well-defined   objects and thus can be represented by 
regularized Feyman diagrams.
It follows from the expression (\ref{C}) of the C-number function 
$C_{\mu}(z,z',x)$, 
which 
was imposed by the necessity of gauge invariance, that the theory keeps trace 
of the 
non-locality in some potentially divergent (1PI) functions through additional 
terms which are regular in the (IR) domain, but whose derivatives are not.
This is the case for the polarization operator and for the vertex function if 
this 
latter is calculated off mass-shell.
Although the polarization operator is transverse, it contains a quadratically 
divergent 
term whose derivative is not regular in the (IR) domain.
Contrary to known gauge invariant regularization procedures, like the Pauli 
-Villars 
regularization where the quadratic divergent piece is not gauge invariant and 
then 
legislated to zero, this term must  be removed by a specific choice of the 
counterterm 
$S_{\Lambda}$.
In the case of massless QED some terms of the non gauge invariant part of 
closed fermions loops are not unambiguously regularized to zero.
This fact is reminiscent of the Nielsen-Ninomiya Theorem for chiral fermions on 
the lattice.
 
Despite the fact that we recover the known result for the 
anomalous magnetic moment of the electron at the one loop order, we must 
investigate 
all 
the implications 
of the (IR) behavior of this regularized form of QED for physical process.
These questions, the problem of non-perturbative renormalization, the 
possibility of 
dynamical mass generation for the photon through a new non-local interaction and 
the 
extension 
of this regularization scheme to non-abelian gauge theories like QCD, where 
non-perturbative effects are known to occur, will be investigated elsewhere.
\newpage
\centerline{\bf Acknowledgememts}
The author would like to thank J. Polonyi for encouragement and useful 
discussions.
\myappendix{Appendix A } 
\appA
The  $ \Gamma_{\mu } $ matrices satisfy :
\beqa
\label {comm}
\  & & \{ \Gamma_{\mu }(r),\Gamma_{\nu }(r) \} ~=~ \{ \Gamma_{\mu 
}(r),\gamma_{\nu } \} ~=~ 
2(g_{\mu \nu}-\frac{r_{\mu } r_{\nu}}{r^2} ) \\
\label {cont}
\ & & r_{\mu } \Gamma_{\mu }(r)~=~0,~~~~~~ \gamma^{\nu } \Gamma_{\mu }(r) 
\gamma_{\nu 
}~=~-2\Gamma_{\mu }(r)  
\eeqa
\myappendix{Appendix B}
\appB
The Fourier transform of $e^{ieL(x,y)}$ is 
\beq
\label {Fpq}
\ F(p,q)~=~\int ~dxdy~ e^{i(px+qy)}e^{ieL(x,y)}.
\eeq
If we define 
\beq
\label {Fnpq}
\ F_n(p,q)~=~\int ~dxdy~ e^{i(px+qy)}L^n(x,y)
\eeq
we have the expansion
\beqa
\label {expanFpq}
\ F(p,q)~&=&~\sum_{n=0}^{+\infty} \frac{(ie)^n}{n!} F_n(p,q) \\
\label {expanFpqO}
\ F_0(p,q)~&=&~((2\pi )^4)^2~\delta (p)\delta (q) ,
\eeqa
and the following recursion formulas hold for the $F_n$'s
\beq
\label {recur}
\ F_{n+1}(p,q)~=~-i\int \dr ~\rho_3(r) \frac{r_{\mu }}{r^2} 
\left(F_n(p-r,q)-F_n(p,q-r)\right)A^{\mu }(r).
\eeq
In addition one can easily show by induction that
\beq
\label {recur2}
\int \dk ~k_{\mu} F_n(p-k,k-p')~=~0,~~~~~~ \int \dk ~  F_n(p-k,k-p')~=~0,
\eeq
for all $p$ and $p'$.
\myappendix{Appendix C}
\appC
If we take as usual the functional derivative of the electron source 
$\eta_a(k)$ 
(\ref{source}) with respect to $\eta_b(k')$, the inverse of the full electron 
propagator 
(\ref{S}) 
\beq
\label {elecone}
\Gamma^{(2)}_{ab}(k,k')~\equiv ~-(2\pi)^4 \frac{\delta^2 \Gamma}{\delta 
\bpsi_a(k) 
\delta 
\psi_b(k')}
\eeq
is defined by
\beq
\label {electwo}
\int dk''~\Gamma^{(2)}_{ac}(k,k'')S_{cb}(k'',k')~=~\delta_{ab}\delta(k-k').
\eeq
In the same manner the inverse of the full photon propagator (\ref{D})
\beq
\label {photone}
\Gamma^{(2)}_{\mu\nu}(k,k')~\equiv ~(2\pi)^4 \frac{\delta^2 \Gamma}{\delta 
A_{\mu}(k) \delta A_{\nu}(-k')}
\eeq
verifies
\beq
\label {phottwo}
\int dk''~\Gamma^{(2)}_{\mu\alpha}(k,k'')D_{\alpha \nu} 
(k'',k')~=~\delta_{\mu\nu}\delta(k-k').
\eeq
The relations (\ref{electwo}) and (\ref{phottwo}) are the basic tools for 
expressing, 
by 
means of Schwinger's technique of functional differentiation, general Green's 
functions 
in 
terms of (1PI) functions and propagators.
For instance, taking the functional derivative of (\ref{electwo}) with respect 
to the 
C-number function $A^{\mu}(-r')$, we obtain the relation
\beq
\label {electhree}
\frac{\delta S_{ba}(p,k)}{\delta A^{\mu}(-r')}~=~-\int 
dk'dk''~S_{bc}(p,k'')\frac{\delta \Gamma^{(2)}_{cd}(k'',k')}{\delta 
A^{\mu}(-r')}S_{da}(k',k).
\eeq
\myappendix{Appendix D}
\appD  
In this appendix we give the asymptotic form of the integrals which we 
encounter in the 
calculation of the relevant (1PI) functions.
We start with the evaluation of the integral
\beq
\label {inone}
\  I~=~ \int dk~\frac{e^{\epsilon (ak^2+2bk)} }{(-k^2-2qk+C)^{\alpha }}.
\eeq
where $ \epsilon $ is an infinitesimal parameter.
At first we make the shift of variable $ k \rightarrow k-q $ and we express the 
denominator of 
the integrand of (\ref{inone}) by means of Schwinger's parametric integral.
After a Wick rotation the integral over $ dk $ is easily performed \cite{CO}, 
and 
we get
\beq
\label {intwo}
\ I~=~i\frac{\pi^2}{\Gamma (\alpha )} e^{\epsilon (aq^2-2bq)} 
\int_{0}^{+\infty} 
du~u^{\alpha  
-1 }e^{-u(q^2+C)} \frac{1}{(a\epsilon+u)^2}e^{-\epsilon^2 \frac{(b-aq)^2}{a 
\epsilon +u 
} 
}
\eeq
Now if we expand $ \exp { [-\epsilon^2 (b-aq)^2/(a \epsilon +u )] } $ in power 
series and 
integrate 
over $ du $ we obtain the integral $I$ in terms of the degenerate 
hypergeometric 
function $ 
\Psi( \alpha ,~ n,~z) $ \cite{GR} as
\beqa
\label {inthree}
\ I~&=&~i\pi^2 e^{\epsilon (aq^2-2bq)} (a\epsilon )^{\alpha -2 } ~ \Bigg \{ 
\Psi\bigg( \alpha 
,~ \alpha -1,~(q^2+C)a \epsilon \bigg)~+~ \sum_{\nu = 1}^{+\infty} 
\frac{(-1)^{\nu }}{\nu !} 
\epsilon^{\nu } (b-aq)^{2\nu } a^{-\nu } \nonumber \\
& & \Psi\bigg( \alpha ,~\alpha -1- \nu ,~(q^2+C)a \epsilon \bigg) \Bigg \} .
\eeqa
When $ n $ is an integer, the degenerate hypergeometric function $ 
\Psi( \alpha ,~ n,~z) $ can be expanded as \cite{NI}
\beqa
\label {infour}
\ \Psi( \alpha ,~ n ,~z)~&=&~ \frac{(-1)^{n+1} }{\Gamma (\alpha -n+1) } \bigg 
\{ 
\sum_{k = 
1}^{n-1} \frac{(-1)^k (k-1)!}{(n-k-1)!} \frac{\Gamma (\alpha -k)}{\Gamma 
(\alpha 
)} 
z^{-k}~+~\sum_{k = 0}^{+\infty} \frac{\Gamma (\alpha +k)}{\Gamma (\alpha )} 
\frac{z^k}{(n+k-1)!k!} \nonumber \\
&&[ \psi(k+1)+\psi(n+k)-\psi( \alpha +k)- \log z ] \bigg \},~~~~~n\geq 1
\eeqa
with the convention that for $n=1$ the first sum is identically zero.
In addition the relation \cite{NI}
\beq
\label {infive}
\ \Psi( \alpha ,~ n ,~z)~=~z^{1-n}\Psi( \alpha -n+1,~2-n,~z)
\eeq
holds.
In the above formulas, $ \psi(x) $ is the $ \psi$ function \cite{GR}.
For $ \alpha =2 $,  we use (\ref{infour}) and (\ref{infive}) and the integral 
(\ref{inone}) 
becomes
\beq
\label {intzero}
\ \int dk~\frac{e^{\epsilon (ak^2+2bk)} }{(-k^2-2qk+C)^2}~=~i\pi^2 \bigg(- \log 
a\epsilon + \psi(1)-1- \log (q^2+C ) + \mathcal{O} (\epsilon ) \log \epsilon 
\bigg) .
\eeq
Similarly for the following integrals we obtain the final result
\beqa
\label {intone}
\ \int dk~\frac{k_{\mu }e^{\epsilon (ak^2+2bk)} }{(-k^2-2qk+C)^2}&=&i\pi^2 
\Bigg[ q_{\mu } 
\log a\epsilon -\frac{1}{2} b_{\mu } a^{-1} - q_{\mu } \bigg( \psi(1)- 
\frac{3}{2}  
-\log (q^2+C) \bigg) \nonumber \\
& & + \mathcal{O} (\epsilon ) \log \epsilon \Bigg] 
\eeqa
\beqa
\label {inttwo}
\ \int dk~\frac{k_{\mu }k_{\nu }e^{\epsilon (ak^2+2bk)} 
}{(-k^2-2qk+C)^2}&=&i\pi^2 \Bigg[ 
-\frac{1}{4} g_{\mu \nu } (a \epsilon )^{-1} - \bigg( \frac{1}{2} g_{\mu \nu } 
(q^2+C) + 
q_{\mu }q_{\nu } \bigg) \log a\epsilon \nonumber \\
\ & &+ \frac{1}{6} (b-aq)_{\mu }(b-aq)_{\nu }  a^{-2} +\frac{1}{2} g_{\mu \nu } 
(q^2+C) 
\bigg( 
\psi(1) -\frac{1}{2}- \log (q^2+C) \bigg) \nonumber \\
\  & &+ \frac{1}{12} g_{\mu \nu } (b-aq)^2 a^{-2}  + \frac{1}{2} \bigg( 
(b-aq)_{\mu 
}q_{\nu 
} 
+(b-aq)_{\nu }q_{\mu } \bigg) 
a^{-1}\nonumber \\
\ & & +q_{\mu }q_{\nu } \bigg( \psi(1) -1- \log (q^2+C) \bigg) + \mathcal{O} 
(\epsilon 
) 
\log 
\epsilon \Bigg] \\
\label {intthree}
\ \int dk~\frac{k_{\mu }k_{\nu }e^{\epsilon (ak^2+2bk)} 
}{(-k^2-2qk+C)^3}&=&i\frac{\pi^2}{2}\Bigg[ \frac{1}{2}g_{\mu  \nu } \bigg 
(\log a\epsilon - \psi(1)+\frac{3}{2} +\log (q^2+C) \bigg )\nonumber \\
\ & &+\frac{q_{\mu }q_{\nu }}{q^2+C} + \mathcal{O} (\epsilon ) \log \epsilon 
\Bigg ]
\eeqa

\end{document}